\documentclass{article}
\usepackage{blindtext}
\usepackage{multicol}
\usepackage{amssymb}
\usepackage{amsmath}
\usepackage{multicol}
\usepackage{graphicx}
\usepackage{float}
\usepackage{caption}
\usepackage{amsmath}
\usepackage{subfig}
\usepackage{epstopdf}
\usepackage{pst-plot}
\usepackage{pgfplots}

\pgfplotsset{width=10cm,compat=1.9}
\usepgfplotslibrary{external}
\restylefloat{table}
\graphicspath{ {./images/} }
\makeatletter \@addtoreset{equation}{section}

\def \be{\begin{equation}}
\def \ee{\end{equation}}
\def \bea{\begin{eqnarray}}
\def \eea{\end{eqnarray}}

\newcommand{\nc}{\newcommand}
\nc{\al}{\alpha} \nc{\bib}{\bibitem} \nc{\la}{\lambda}
\nc{\C}{\mbox{\hspace{1.24mm}\rule{0.2mm}{2.5mm}\hspace{-2.7mm} C}}
\nc{\R}{\mbox{\hspace{.04mm}\rule{0.2mm}{2.8mm}\hspace{-1.5mm} R}}
\graphicspath{ {./Holographic f(Q,T)/} }
\begin{document}

\title{Dynamics of Linear Scalar Perturbation in f(Q)+f(T) class of F(Q,T) Gravity }
\author{H.Filali$^{1}$, R.Ahl Laamara$^{1,2}$, M. Bennai$^{2}$,E.H.Saidi$^{1}$\thanks{%
Corresponding authors: houda\_filali3@um5.ac.ma \\
pr.
 mdbennai@yahoo.fr, mohamed.bennai@univh2c.ma \\
 } \\
$^{1}${\small Lab of High Energy Physics, Modeling and Simulations,}\\
\ {\small Faculty of Science, Mohammed V University in Rabat, Morocco}\\
$^{2}${\small CPM, Center of Physics and Mathematics}\\
\ {\small , Faculty of Science, Mohammed V University in Rabat, Rabat,Morocco}}

\maketitle

\begin{abstract}
In this work, we study the $f(Q,T)$ model of symmetric teleparallel modified gravity in the framework of cosmological perturbation theory. Using a general approach, we extract the differential matter density equation then we simplify it as a second-order equation by considering the sub-Hubble approximation. Our analysis is then based on two different forms of $f(Q,T)$ that we study in a classic approach and again using Holographic dark energy. Our initial results yield a significant divergence from the perturbed behavior of the $\Lambda CDM$ model, imposing stringent constraints on the feasibility of this class of theories but the HDE contribution triggers an interesting discussion.

\textbf{Keywords:} Modified gravity, Symmetric Teleparallel gravity, Scalar Perturbation, Holographic dark energy.

\end{abstract} 

\section{\protect\bigskip Introduction}

In Einstein's model of General Relativity, the gravitational interactions of the universe are the result of curvature taking place in the geometry of space-time. That curvature is the result of the metric tensor and Levi-Civita connection. That approach, while being regarded as the best suited to describe the dynamics of the cosmos according to observation, comes with a number of challenges. In this regard, alternative models of modified gravity were proposed throughout the literature in hopes of achieving a more general solution for gravitation. One of the most interesting alternatives of recent years is the class of Teleparallel gravities \cite{1,2} which rely on the principle of a curvature-free space-time $(R^{\alpha}_{\beta \mu \nu}=0)$.
In the realm of metric affine geometries, we can distinguish two principal models, Teleparallel gravity, with torsion and vanishing curvature \cite{2} and Symmetric Teleparallel gravity, for which both torsion and curvature vanish and the gravity is solely described by the non-metricity $Q$ \cite{3,4}.This last case is what we will be investigating in this paper. Furthermore, GR can be reconstructed in both teleparallel and symmetric teleparallel models of gravity, giving rise to functioning equivalent theories \cite{1,6,7}. \\ 
The $F(Q)$ class of gravity theories rely on a different geometrical representation of spacetime that is based on the variation of the magnitude of vectors during parallel transport. This approach proved to be of big interest, gathering a lot of attention in recent years which was even more accentuated after promising initial results both in the geometry and physics of the model through gravitational waves tests, energy conditions, and observational constraints \cite{8,9,10}.
From a cosmological perspective, studies demonstrated that the universe's accelerated expansion is a natural feature of the geometry of the $f(Q)$ model, not requiring the addition of a dark energy element \cite{11}. 
\\ 

The question of perturbation in metric-affine geometries has been extensively explored in the literature \cite{38,39,40,41}. As it is the case for $\Lambda$CDM or other gravity theories, an analysis of the cosmological perturbation behavior of the gravity model is essential to better assess the extent of its applicability in a realistic scenario. Additionally, the study of perturbation for different metric-affine geometries shows important discrepancies depending on whether the models include dynamical fields or exhibit a special geometrical behavior. These investigations are conducted in the framework of both metric and flat teleparallel connection perturbation, exploring the dynamics of gauge-invariant perturbation in the general class of teleparallel gravity with non-vanishing torsion, non-minimal coupling to scalar fields, including both flat and curved backgrounds. 

In the present work, we use the $f(Q,T)$ model of gravity \cite{12}, an extension of symmetric teleparallel gravity, as a function of the non-metricity $Q$ which describes the geometry of space-time and the trace of the stress-energy tensor $T$, representing the matter content of the universe. This function yields a coupling between $Q$ and $T$ that is not minimal, resulting in a divergence and the appearance of an extra force which interpretation will be discussed later in the paper. 
The non-minimal coupling also leads to the non-conservation of the stress-energy tensor, as has been observed in previous papers \cite{12,7}, resulting in the energy transfer between the fabric of the universe and its content. This phenomenon proved to have some impact on the high energy level in the realm of particle physics and quantum fluctuations. Further investigation in this direction could be interesting \cite{12,13}. 

 A description of the cosmological evolution of the $f(Q, T)$ model through late-time expansion using a Holographic dark energy density was obtained in a previous work,  which had very similar behavior to $\Lambda CDM$ at present times, while also being compatible with energy conditions and cosmographic parameters\cite{10,14}.
These results motivated the study of $f(Q,T)$ beyond the framework of background cosmology, using the theory of linear cosmological perturbation to better understand and examine our modified gravity model's behavior in a realistic environment and properly analyze the density contrast growth.
\\ \\
For the purpose of this communication, we obtain the exact equations for the matter density evolution to establish the dynamics of linear perturbations for $f(Q) + f(T)$ type gravitational action. While presenting interesting features,  this class of the model has not been thoroughly explored yet. A first attempt at investigating linear perturbation in $f(Q,T)$ was introduced in \cite{16}, but the explicit numerical results were not obtained, which what this paper aims to do. An additional motivation for this work is the results obtained in \cite{7} for a similar class of $f(R,T)$ gravity, for which the matter density evolution was investigated. The unusual behavior of that model in the perturbed framework is what prompted us to initiate a similar study in the $f(Q,T)$ model.

In order to facilitate our approach, we mimic the quasi-static approximation by considering sub-Hubble modes as initial value allowing us to neglect the time derivatives of Bardeen's potentials, preserving density perturbations only. We then analyze the $f(Q,T)$ model matter density evolution for the form $f(Q,T)=\alpha Q^{n+1}+\beta T$, a general form that is compatible with the constraints of our approach. We investigate this same cases again in the framework of Holographic dark energy in order to establish the impact of quantum fluctuations on the perturbed behavior.
  \\ 

This paper is organized in the following manner: \textbf{Sec. 2}  The background cosmological equations for the $f(Q,T) = f(Q) + f(T)$ model are introduced and the divergence equation for the non-conservation of the stress-energy tensor is established, \textbf{Sec. 3} the scalar perturbation equations for $f(Q,T)$ are derived and the differential equation for the matter density is extracted; In \textbf{Sec. 4} the evolution of the density contrast is investigated in the framework of the $f(Q,T)$ model and tested against GR both in the original and again with HDE.\textbf{ Sec. 5} is for conclusion and discussion. 
\\ \\

\section{ General model of $F(Q,T)$ in Flat Universe}

The symmetric teleparallel gravity is defined by a vanishing Levi-Civita connexion \cite{26} and the non-metricity tensor
 $Q_{\gamma\mu\nu}=\nabla_{\gamma}g_{\mu\nu}$. We can derive the trace of the non-metricity tensor following the procedure :

\begin{equation} Q_{\delta}=g^{\mu\nu}Q_{\delta\mu\nu} , \tilde{Q}_{\delta}=g^{\mu\nu}Q_{\mu\delta\nu}\end{equation}

The Superpotential is then obtained by the following equation,
\begin{equation} P^{\delta}_{\mu\nu}=-\frac{1}{2}Q^{\delta}_{\mu\nu}+\frac{1}{4}(Q^{\delta}-\tilde{Q}^{\delta})g_{\mu\nu}-\frac{1}{4}\delta^{\delta}(\mu Q\nu) \end{equation}

Which allows us to define the non-metricity scalar $Q$ as, 
\begin{align}
   & Q=-Q_{\delta\mu\nu}P^{\delta\mu\nu} \\ \notag
   &  = \frac{-1}{4} \left(Q^{\delta\nu\lambda}Q_{\delta\nu\lambda}+2Q^{\delta\nu\lambda}Q^{\lambda\delta\nu}-2Q^{\lambda}\bar{Q}^{\lambda}+Q^{\lambda}Q_{\lambda} \right) \\ \notag
\end{align}

 Now that we have established the essential parameters of our model, we write the action for our $f(Q,T)$ gravity as follow \cite{14,26} : 

\begin{equation}
S=\int[\frac{1}{16\pi}f(Q,T)+L_{m}]\sqrt{-g}d^{4}x
\label{a1}
\end{equation}%

Where $f$ is an arbitrary function of $Q$ and $T$, the trace of the stress-energy tensor. $L_{m}$ is the matter Lagrangian and $g$ is the determinant of the metric tensor $g_{\mu \nu }$. \\ 

The stress-energy tensor is usually defined as  : \begin{equation}T_{\mu\nu}=\frac{-2}{\sqrt{-g}}\frac{\delta\sqrt{-g}L_{m}}{\delta g^{\mu\nu}}\end{equation}

And we obtain our gravity model's field equation by varying the action with respect to the metric $g_{\mu \nu}$ and equating it to zero :
\begin{equation}
\begin{gathered}
-\frac{2}{\sqrt{-g}}\nabla_{\delta}(f_{Q}\sqrt{-g}P^{\delta}_{\mu\nu})-\frac{1}{2}fg_{\mu\nu}+f_{T}(T_{\mu\nu}+\theta_{\mu\nu}) \\ -f_{Q}(P_{\mu\delta\alpha}Q^{\delta\alpha}_{\nu}-2Q^{\delta\alpha}_{\mu}P_{\delta\alpha\nu}) =8\pi T_{\mu\nu}
\end{gathered}
\end{equation}

where $f_{Q}=\frac{df}{dQ}$, $f_{T}=\frac{df}{dT}$ and $\nabla_{\delta}$ is the covariant derivative. \\

We use the Lagrangian multiplier method, ensuring that the connection is both flat and torsion-free. This allows us to find the variation of the metric with respect to the connection by introducing $\nabla _{\mu}\nabla_{\nu}$ into the action, eliminating the Lagrangian multiplier coefficients. This is possible due to $\mu$ and $\nu$ being anti-symmetric in the coefficients. More detailed calculations can be found in \cite{12}

Variation of the action with respect to the connection then yields the connection field equation : 
\begin{equation}
    \nabla _{\mu}\nabla_{\nu}\left( \sqrt{-g}f_{Q}P^{\mu\nu}_{\alpha}+4\pi H^{\mu\nu}_{\alpha} \right)=0
\end{equation}
With $H^{\mu\nu}_{\alpha}$ being the hypermomentum.

In order to study the cosmological evolution of the $f(Q,T)$ theory, we choose to work in a flat, homogeneous and isotropic FLRW spacetime :
\begin{equation}ds^2=-N(t)^2dt^2+a(t)^2(dx^2+dy^2+dz^2) \end{equation}
 where $a(t)$ is the scale factor and $N(t)=1$ is the lapse function for a flat spacetime. \\

The Hubble parameter is defined as $H\equiv \dot{a}/a$, by considering the standard spherical-Cartasian coordinate transformation. The vanishing of the curvature imposes the connection to be purely inertial, allowing for the connection to completely vanish. The gauge in which we set these parameters as the coordinates is therefor the coincident gauge  with $\Gamma^{\alpha}_{\nu\mu} = 0$ ,we have $Q=6H^{2}$. \cite{42,43,44} \\
 The stress-energy tensor of a perfect fluid is written as  : 
\begin{align}T_{\nu}^{\mu}=diag(-\rho ,p,p,p)
\end{align}
Where $\rho$ and $p$ are the energy density and pressure of the cosmic fluid, respectively. \\
The modified Friedman equations take the following forms
 \begin{equation}
8\pi \rho =\frac{f}{2}-6FH^{2}-\frac{2\tilde{G}}{1+\tilde{G}}(\dot{F}H+F\dot{H}) \end{equation}
and
\begin{equation}8\pi p= -\frac{f}{2}+6FH^{2}+2(\dot{F}H+F\dot{H})
 \end{equation}
Where $F=f_{Q}$ , $8\pi\tilde{G}=f_{T}$ and dot represents derivative with respect to time.
\\

 We can then obtain the evolution of the Hubble function : 
\begin{equation}
\dot{H}+\frac{\dot{F}}{F}H=\frac{4\pi }{F}(1+\tilde{G})(\rho +p) 
\end{equation}
To retrieve a form of the cosmological equations that is equivalent to GR, we consider the effective Friedmann energy density and pressure ; 
\begin{equation} 8\pi \rho _{eff}=3H^{2}=\frac{f}{4F}-\frac{4\pi}{F}((1+\tilde{G})\rho +\tilde{G}_{p})
\end{equation}
and
 \begin{equation}
\begin{gathered}
-8\pi\rho_{eff}=2\dot{H}+3H^{2} \\ =\frac{f}{4F}-\frac{2\dot{F}H}{F}+\frac{4\pi}{F}((1+\tilde{G})\rho +(2+\tilde{G})p)
\end{gathered}
\end{equation}

 It is then possible to obtain the effective conservation equation $\dot{\rho}_{eff}+3H\rho_{eff}=0$, assuming negligible pressure. \\

\subsection{Divergence of the stress-energy tensor}

As we mentioned in the Introduction of this paper, one of the most interesting particularities of the $f(Q,T)$ model is the non-conservation of the stress-energy tensor $T$ due to its coupling to $Q$. It is naturally followed by the apparition of a non-null divergence equation that can be expressed as follow
\begin{eqnarray}
\hspace{-0.5cm} &&\mathcal{D}_{\mu }T_{\ \ \nu }^{\mu }=\frac{1}{f_{T}-8\pi }%
\Bigg[-\mathcal{D}_{\mu }\left( f_{T}\Theta _{\ \ \nu }^{\mu }\right) -\frac{%
16\pi }{\sqrt{-g}}\nabla _{\alpha }\nabla _{\mu }H_{\nu }^{\ \ \alpha \mu } 
\notag \\
\hspace{-0.5cm} &&+8\pi \nabla _{\mu }\bigg(\frac{1}{\sqrt{-g}}\nabla
_{\alpha }H_{\nu }^{\ \ \alpha \mu }\bigg)-2\nabla _{\mu }A_{\ \ \nu }^{\mu
}+\frac{1}{2}f_{T}\partial _{\nu }T\Bigg],
\end{eqnarray}%
where $H_{\gamma }^{\ \ \mu \nu }$ is the hyper-momentum tensor density
defined as,%
\begin{equation}
H_{\gamma }^{\ \ \mu \nu }\equiv \frac{\sqrt{-g}}{16\pi }f_{T}\frac{\delta T%
}{\delta \widetilde{\Gamma }_{\ \ \mu \nu }^{\gamma }}+\frac{\delta \sqrt{-g}%
\mathcal{L}_{M}}{\delta \widetilde{\Gamma }_{\ \ \mu \nu }^{\gamma }}.
\end{equation}

From connection field equation (2.6), we can write (2.16) as ; 
\begin{align}
&-(f_{T}+8\pi)\mathcal{D}_{\mu}T^{\mu}_{\nu}+f_{T}\partial_{\nu}p-\frac{1}{2}f_{T}\partial_{\nu}T=\frac{1}{\sqrt{-g}}Q_{\mu}\nabla _{\alpha}(\sqrt{-g}f_{Q}P^{\alpha\mu}_{\nu}) \\
& +\frac{2}{\sqrt{-g}}\nabla _{\mu}\nabla _{\alpha}(\sqrt{-g}f_{Q}P^{\mu\alpha}_{\nu})\equiv B_{\nu} \notag
\end{align}
Since  $\hspace{0.2cm}%
\mathcal{D}_{\mu }T_{\ \ \nu }^{\mu }\neq 0$, we can deduce that the stress-energy tensor is not conserved and leads to a divergence. From a phenomenological perspective, the appearance of an extra parameter in the equation indicates that our gravitational model is not conserved and an exchange of energy takes place in our system leading to particle production with $B_{\nu}$ being the amount of energy created or annihilated. This extra energy, which manifests as an additional force, can then be interpreted as either an exotic field or quantum fluctuations. In some cases, the energy transfer can even take place between the dark components of the universe. The full impact of this divergence will be discussed in the following sections.

\section{Scalar Perturbations in $F(Q,T)$ Gravity}

In the previous section, we established the field equations of the $f(Q,T)$ model and divergence of the stress-energy tensor allowing us to describe the cosmological evolution of the aforementioned gravity theory at background level. But for any gravity model, a full comprehension of the viability of the system can only be obtained through an investigation of its behavior in the framework of cosmological perturbations.

We consider linear cosmological perturbation theory in the framework of $f(Q,T)$ gravity as introduced in \cite{1} and take $h_{\mu\nu}$ as the first order perturbation, which is presented in the following form \cite{43, 44}: 
\begin{equation}
    g_{\mu\nu}\to g_{\mu\nu}+\delta g_{\mu\nu}
\end{equation}
\begin{equation}
    g_{\mu\nu}=a^{2}\left( \eta_{\mu\nu}+h_{\mu\nu} \right)
\end{equation}
\\
For the perturbation connection components, we introduced the perturbed connection following the same process as the metric : 
    $\Gamma^{\alpha}_{\mu\nu}\to \Gamma^{\alpha}_{\mu\nu}+\delta \Gamma^{\alpha}_{\mu\nu}$ ; 
    with $ \delta \Gamma^{\alpha}_{\mu\nu} = \nabla _{\mu}\delta\Lambda^{\alpha}_{\nu}$

    When it comes to symmetric-teleparallel theories we find $\delta\Lambda^{\mu}_{\nu} = \nabla _{\nu}\delta\chi^{\mu}$. We can thus always use coordinate transformation to set the affine connection perturbations to zero $(\delta\chi^{\mu}\to 0)$. This allows us to use a gauge equivalent to the coincidence gauge, making the connection vanish for any flat, torsion-free connection \cite{43}. We lose the diffeomorphism invariance of the equations, and we are therefore not working in the usual gauge invariance of cosmological perturbation. A detailed analysis of behavior under diffeomorphism transformation, including gauge transformations, was provided in \cite{51}. 
    
We choose to work with scalar perturbations because they are the only components of the perturbed metric whose instability parameters grow with time, and this paper is particularly focused on manifestations of perturbation during late time expansion. \\
We then find the perturbed components of the stress-energy
tensor in this gauge \cite{1} ; 
\begin{equation}
    T^{0}_{0}=-\bar{\rho}-\delta\rho ,\:\: T^{i}_{j}=\left( \bar{p}+\delta p \right)\delta_{ij} ,\:\: T^{0}_{i}=\left( -\bar{\rho}+\bar{p} \right)\left( \partial_{i}(\upsilon-W)+(\upsilon_{i}-W_{i}) \right)\:\: and \:\: T^{i}_{0}=\left( \bar{\rho}+\bar{p} \right)\left( \partial_{i}\upsilon+\upsilon_{i} \right)
\end{equation}
The bar over the density and pressure refers to these parameters at background value and  $v$ represents the velocity perturbation potential. \\

Using the field equations we found (2.7) with the scalar part of the perturbed metric $h_{\mu \nu}$ (3.1) and taking the $00$ component, we can extract the first order perturbed equation 
\begin{align}
\label{eqn:density}
& a^2 \delta \rho \left( 8\pi + \frac{3}{2} \bar{f}_T - \bar{f}_{TT} (\bar{\rho} + \bar{p}) \right) + a^2 \delta p \left( 3\bar{f}_{TT} (\bar{\rho} + \bar{p}) - \frac{1}{2} \bar{f}_T \right) \nonumber \\
&= 6\left( \bar{f}_Q + 12 \frac{\bar{f}_{QQ}}{a^2} \mathcal{H}^2 \right) \mathcal{H} \left(\mathcal{H}\Psi + {\Phi}^{'}\right) + 2\bar{f}_Q k^2 \Phi -2\left( \bar{f}_Q + 3 \frac{\bar{f}_{QQ}}{a^2} \left( {\mathcal{H}}^{'} + \mathcal{H}^2 \right) \right) \mathcal{H} k^2 W.
\end{align}
\\ \\

To obtain the equation for the velocity potential, we take the perturbed component of the stress-energy tensor $T^0_{\;\;i}$ :

\begin{align}
\label{eqn:velocityPerturbation}
    v &= \frac{1}{(\bar{f}_T + 8\pi)(\bar{\rho}+\bar{p})a^2} \biggl[ 2 \left(\bar{f}_Q+3\frac{\bar{f}_{QQ}}{a^2} (\mathcal{H}'-\mathcal{H}^2)\right) \mathcal{H} \Psi + 12 \frac{\bar{f}_{QQ}}{a^2} \mathcal{H}^2 \phi' \nonumber \\
    &+ 18 \mathcal{H} \frac{\bar{f}_{QQ}}{a^2} (\mathcal{H}'-\mathcal{H}^2) \phi + 2\bar{f}_Q \Phi' - 4k^2 \frac{\bar{f}_{QQ}}{a^2} \mathcal{H}^2 W \biggr]
\end{align}
We are most interested in the matter density equation considering that the growth evolution of that parameter is intricately dependent on the theory of gravity that it's based on. Moreover, it is known that cosmological perturbations are directly related to the information imprinted on the CMB and the matter power spectrum of galaxy clusters, the growth of matter density would then be a strong indicator of the viability of these theories. 

If we want to find the equation for $\delta = \dfrac{\delta \rho}{\bar{\rho}}$, we take equations (3.3) and (3.4), which are the Continuity and Euler equations, and we extract the perturbed time-time component of the divergence equation 
\begin{align}
    \label{eqn:continuityEquation}
    &\left[ \bar{f}_T + 8\pi + \frac{1}{2} \bar{f}_T \left( 1 -  \frac{\delta p'}{\delta \rho'} \right) \right] \delta' - \biggl[ 3 \mathcal{H} (\bar{f}_T + 8\pi) \left( \frac{(\bar{f}_T + 8\pi) w - \frac{1}{2} \bar{f}_T (1-c_s^2)}{\bar{f}_T + 8\pi + \frac{1}{2} \bar{f}_T (1-c_s^2)} - \frac{\delta p}{\delta \rho} \right) \nonumber \\
    &+ \frac{3\mathcal{H} (\bar{f}_T + 8\pi) (1+w)}{2\left(\bar{f}_T + 8\pi + \frac{1}{2} \bar{f}_T (1-c_s^2)\right)} \left( \bar{f}_T \left( 1 - \frac{\delta p'}{\delta \rho'} \right) + \bar{f}_{TT} \bar{\rho} \left( 1 - 3 \frac{\delta p}{\delta \rho} \right) (1-3 c_s^2) \right) \nonumber \\
    &+ 3\mathcal{H} \bar{f}_{TT} \bar{\rho} \left( 1 - 3 \frac{\delta p}{\delta \rho} \right) (1+w) \left( \frac{\bar{f}_T (1-c_s^2)}{\bar{f}_T (3-c_s^2) + 16\pi} \right)  \biggr] \delta - (\bar{f}_T + 8\pi ) \biggl[ (1+w) k^2 (v-W) \nonumber \\
    &+ 3\phi' (1+w) \biggr] = \frac{\delta B_0}{\bar{\rho}},
\end{align}
We now consider the spatial components of the perturbed divergence and find the Euler equation
 
\begin{align}
\label{eqn:EulerEquation}
    v' + \frac{3\bar{f}_T + 8\pi - c_s^2 (8\pi + 5\bar{f}_T)}{\bar{f}_T + 8\pi + \frac{1}{2} \bar{f}_T (1-c_s^2)} \mathcal{H} v &+ \frac{\delta p}{\bar{\rho} + \bar{p}} + \Psi - \frac{\bar{f}_T}{(\bar{f}_T + 8\pi) \mathcal{H} (\bar{\rho} + \bar{p})} \delta p \nonumber \\
    &= \frac{i k_j \delta B_j}{(\bar{f}_T + 8\pi) (\bar{\rho}+\bar{p})k^2},
\end{align}

The result is two differential equations of $\delta$ and $v$ that are linked to each other. Considering that an analytical solution for (3.5) would require long and tedious calculations, an alternative approach was proposed in \cite{1}, where we consider a non-matter dominated era resulting in negligible pressure and $w = 0 = c_s^2$, eqs. (3.5) and (3.6) become

\begin{align}
\label{eqn:continuityEquationDust}
    \left( \frac{3}{2} \bar{f}_T + 8\pi \right) \delta' &+ \frac{3\mathcal{H}}{3\bar{f}_T + 16\pi} \left[ (\bar{f}_T + 8\pi) \bar{f}_T - (\bar{f}_T + 8\pi) (\bar{f}_T + \bar{f}_{TT} \bar{\rho}) - \bar{f}_{TT} \bar{\rho} \bar{f}_T \right] \delta \nonumber \\
    & - (\bar{f}_T + 8\pi) \left[ k^2(v-W) + 3\phi' \right] = \frac{\delta B_0}{\bar{\rho}}
\end{align}
\begin{align}
\label{eqn:EulerEquationDust}
    v' + \frac{6\bar{f}_T + 16\pi}{3\bar{f}_T + 16\pi} \mathcal{H} v + \Psi = \frac{i k_j \delta B_j}{(\bar{f}_T + 8\pi) \bar{\rho} k^2}.
\end{align}

To further simply our operation, still supposing our matter growth evolution in the radiation era, we find the following Bardeen's potentials by neglecting the time derivative :  

\begin{equation}
    \Phi = \frac{a^2 \delta \bar{\rho} \left( 8\pi + \dfrac{3}{2} \bar{f}_T - \bar{f}_{TT} \bar{\rho}\right)}{2 \bar{f}_Q k^2}.
\end{equation}

 \begin{equation}
      \Psi = \frac{a^2 \delta \bar{\rho} \left( 8\pi - \bar{f}_{TT} \bar{\rho} \right)}{2 \bar{f}_Q k^2}
 \end{equation}

We note that by setting $f_{T}$ to zero we find $\Psi = \Phi$ and we also recover the Bardeen potentials for the perturbed $f(Q)$ gravity as seen in \cite{37}.

By considering the sub-Hubble mode $\mathcal{H}<<k$ and negligible time derivative of the eq. (3.4) we are able to recover the Continuity and Euler equations as follow

\begin{align}
   & \left( \frac{3}{2} \bar{f}_T + 8\pi \right) \delta' + \frac{3\mathcal{H}}{3\bar{f}_T + 16\pi} \left[ (\bar{f}_T + 8\pi) \bar{f}_T - (\bar{f}_T + 8\pi) (\bar{f}_T + \bar{f}_{TT} \bar{\rho}) - \bar{f}_{TT} \bar{\rho} \bar{f}_T \right] \delta \nonumber \\
   & - \frac{(\bar{f}_T + 8\pi)}{2} \left( k^2  + \frac{(\bar{f}_T+8\pi)\bar{\rho}a^4}{2\bar{f}_{QQ} \mathcal{H}^2} \right) v = 0,
\end{align}
and
\begin{align}
    v' + \frac{6\bar{f}_T + 16\pi}{3\bar{f}_T + 16\pi} \mathcal{H} v + \frac{a^2 \bar{\rho} (8\pi - \bar{f}_{TT} \bar{\rho})}{2\bar{f}_Q k^2} \delta + \frac{1}{2} \left( \frac{\bar{f}'_{QQ}}{\bar{f}_{QQ}} - 3\left(\frac{\mathcal{H}'}{\mathcal{H}}+\mathcal{H}\right) \right) v = 0.
\end{align}

We note that while assuming a negligible time derivative for the Bardeen potentials and perturbed velocity, and considering the sub-Hubble mode, these equations are not representative of a quasi-static approximation of the $f(Q,T)$ theory as this limit has not been thoroughly explored yet. 

Therefore, we have two equations that will allow us to find a numerical solution for the matter growth density $\delta$. 

\section{Results and Discussion}

In the following section, we will be combining the equations (3.11) and (3.12) to find the expression for the density contrast equation in our model of $f(Q,T)$ gravity yielding : 
\begin{align}
    & \left(\frac{3}{2} f_{T}+8\pi \right) \delta^{''}+\Biggl\{\left[\frac{\left(6f_{T}+16\pi \right)}{\left(3f_{T}+16\pi \right)})\mathcal{H}+\frac{1}{2} \left(\frac{f'_{QQ}}{f_{QQ}} -3(\frac{\mathcal{H'}}{\mathcal{H}}+\mathcal{H})\right)\right]](\frac{3}{2} f_{T}+8\pi) \notag \\
    & +\frac{3\mathcal{H}}{(3f_{T}+16\pi)} \left[(f_{T}+8\pi)f_{T}-(f_{T}+8\pi)(f_{T}+f_{TT}\rho)-f_{TT}\rho f_{T}\right] \Biggl\}\delta^{'}  \\
    & +\Biggl\{\frac{3\mathcal{H}}{(3f_{T}+16\pi)} \left[\frac{(6f_{T}+16\pi)}{(3f_{T}+16\pi)}\mathcal{H}+\frac{1}{2} \left(\frac{f'_{QQ}}{f_{QQ}} -3(\frac{\mathcal{H'}}{\mathcal{H}}+\mathcal{H})\right)\right][(f_{T}+8\pi) f_{T}-(f_{T}+8\pi)(f_{T}+f_{TT}\rho)-f_{TT}\rho f_{T} ] \notag \\
    & +\left(\frac{(f_T+8\pi)}{2}\right) \left(k^{2}+\frac{(f_{T}+8\pi)\rho a^{4}}{2f_{QQ} \mathcal{H}^{2}}\right)\left(\frac{a^{2}\rho(8\pi-f_{TT} \rho)}{2f_{Q} k^{2}}\right)\Biggl\}\delta = 0 \notag
\end{align}
In addition, for GR, the quasi-static equation of the mass density $\delta$ is
\begin{equation}
    \delta ''+ H\delta'-4\pi G\rho_{0}a^{2}\delta=0
\end{equation}

Therefore, we obtain the complete set of equations that allow us to analyze the evolution of scalar linear perturbations for our model, and compare it with the known behavior of the perturbed $\Lambda CDM$. \\
First thing we notice with equation (4.1) is the appearance of an extra $f_{T}$ dependent term in all the $\delta$ and its time derivative coefficients and secondly, we see the appearance of $k^2$ modes in the $\delta$ coefficient which, while is present in the perturbed scalar equations of $f(Q)$ models, is not part of the GR equation. This will allow us to investigate the impact $k$ modes have when correlated with non-minimal coupling of gravity with matter, especially at late-time when the $k^{2}$ is usually dominant for sub-Hubble modes.
\\
On the other hand, we can see how equation (4.1) has a substantial length making it difficult to compute. Following the same simplification measures we used earlier, we determine that $k^{2} = M_{P}^{2} =(10^{19} GeV)^{-2}$ and $f_{T} = \rho^{-1/2}_{critical} = (10^{-3} eV)^{-2}$, equation (4.1) becomes 
\begin{align}
    &\frac{3}{2} f_{T} \delta^{''}+\Biggl\{\left[2f_{T}\mathcal{H}+\frac{1}{2} \left(\frac{f'_{QQ}}{f_{QQ}} -3(\frac{\mathcal{H'}}{\mathcal{H}}+\mathcal{H})\right)\right]\frac{3}{2} f_{T} +\frac{\mathcal{H}}{f_{T}} \left[f_{T}^{2}-f_{T}(f_{T}+f_{TT}\rho)-f_{TT}\rho f_{T}\right] \Biggl\}\delta^{'} \notag \\
   & +\Biggl\{\frac{3}{2} f_{T}+\frac{\mathcal{H}}{f_{T}}\left[2 f_{T}\mathcal{H}+\frac{1}{2} \left(\frac{f'_{QQ}}{f_{QQ}} -3(\frac{\mathcal{H'}}{\mathcal{H}}+\mathcal{H})\right)\right][f_{T}^{2}-f_{T}(f_{T}+f_{TT}\rho)-f_{TT}\rho f_{T} ] \\
    &+\frac{f_{T}}{2}\left(k^{2}+\frac{f_{T}\rho a^{4}}{2f_{QQ} \mathcal{H}^{2}}\right)\left(\frac{a^{2}\rho(8\pi-f_{TT} \rho)}{2f_{Q} k^{2}}\right)\Biggl\}\delta=0 \notag
\end{align}

It follows that the final result is a differential equation that can only be solved numerically.

For that purpose, we will be choosing two different models of the $f(Q,T)$ gravity that clearly transcribe the impact of each parameter in the equation as it evolves in time.

\subsection{ $f(Q,T)=\alpha Q^{n+1}+\beta T$}

We choose for this work to investigate the standard form of the $f(Q,T)$ gravity that has been explored many times in the literature. Moreover, we use this form of the function $f(Q,T)=\alpha Q^{n+1}+\beta T$ rather than its simpler versions in order to avoid singularity at critical points such as in $\left(k^{2}+\frac{f_{T}\rho a^{4}}{2f_{QQ} \mathcal{H}^{2}}\right)$ and $\left(\frac{a^{2}\rho(8\pi-f_{TT} \rho)}{2f_{Q} k^{2}}\right)$.
Which leads us to the following result : 
\begin{align}
    & \frac{3}{2} \beta \delta^{''}+\Biggl\{2\beta \mathcal{H}+\frac{3}{4} \beta(6\mathcal{H}^{2(n-1)} \mathcal{H}^{'(n-1)}-\frac{3\mathcal{H}'}{\mathcal{H}}-3\mathcal{H})\Biggl\} \delta' \\
   &  +\Biggl\{\frac{3}{2}\beta+\frac{\beta}{2} (k^{2}+\frac{\beta\rho a^{4}}{12\alpha n(n+1)\mathcal{H}^{2n}})(\frac{(a^{2} \rho 8\pi)}{16\mathcal{H}^{2}k^{2}})\Biggl\}\delta=0 \notag \\ \notag
\end{align}
We parametrize the constant values $\alpha$, $\beta$ and $n\neq 1$ according to observational constraints \cite{10,14}, as you'll notice, if we set it to $n=0$ then we retrieve the standard $f(Q,T)$ form which is $f(Q,T)= \alpha Q + \beta T$. We can then consider the form we used for our model as a generalization encompassing various cases.  \\
Since we are still supposing sub-Hubble modes at high redshift, $\delta$ is assumed to be proportional to the scale factor at conformal time.  \\
In a second step, we introduce the growth factor $f(z)$, which is also essential the investigate the behavior of the matter density in a way that makes it possible to test against observational data.
The growth factor is presented as \cite{45, 46} : 
\begin{equation}
    f \equiv \frac{dln\delta_{m}}{lna}=-(1+z)\frac{\delta'_{m}(z)}{\delta_{m}(z)}
\end{equation}
In Fig. 1 we have plotted the evolution of the density contrast for the $\Lambda CDM$ model as a reference and the evolution of the same density contrast in our $f(Q,T)$ model for several modes with respect to the redshift z.
As we know, sub-horizon perturbations are stabilized during radiation domination by pressure, the slight acceleration that occurs during the matter-dominated era will substantially affect the density contrast while the amount of matter in the universe remains the same. \\

In our case, one can see how, while the behavior for all models is the same during the radiation era, the $k$ modes take over equation (4.6) during late-time evolution and the behavior of the density contrast for $f(Q,T)$ is completely different from the GR model when we hit the point of expansion around z=2. At the same time, GR evolves in an almost power-law fashion.
We mainly notice that while oscillations for the $\Lambda CDM$ model are not null, they are heavily damped when compared to the strong oscillations for our $f(Q,T)$ model between z=0 and z=1.7 for all three modes.
\\
When comparing these results with the ones in \cite{19}, where a similar approach was used for the $f(R,T)$ model, we can see a similar but much more damped oscillation behavior for the evolution of $\delta$ through time, prompting the question for the implication of the stress-energy tensor $T$ and how it impacts the evolution of the matter density, seeing that oscillations occur in both models where $T$ is present although being two completely different forms of gravity. Especially when we know that the perturbed evolution of $f(Q)$ ranges in both positive and negative values depending on the choice of $\alpha$ but does not oscillate \cite{37}.
This result indicates that the oscillations are a direct consequence of the amplitude evolution of baryonic matter, which might be sensitive to the presence of scalar fields simulating a frequency resonance at late time. The short-term splash of amplitude can be interpreted as a perturbative manifestation of an undetermined matter density presence. 
\\
In Fig. 2 we plotted the growth factor $f(z)$ and as observed, the results obtained by our model show a clear divergence from LCDM but also from some cosmographic constraints \cite{52}. Therefore, this could be interpreted as a clear indication that the model is not effective and is not a viable alternative for modified gravity, as the increasing growth factor is incompatible with expectations set by LCDM. On the other hand, reconstruction from data collected via BAO + SN has shown conflicting results for the present time growth index at varying confidence \cite{53}. Other works have demonstrated different results for different models when constrained by data \cite{54,55}. This prompts us to acknowledge that while the most likely explanation for our results is that our model does not work, further investigation into future observational data is necessary to establish a final answer.

\begin{figure}
    \centering
    \includegraphics[width=0.75\linewidth]{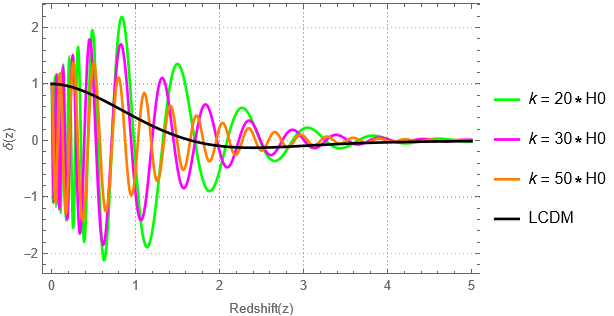}
    \caption{Evolution of density contrast with respect to redshift (z) for the case $f(Q,T)=\alpha Q^{n+1}+\beta T$ with $\alpha = -2.5$ , $\beta = -11.5$ and $n= 1$ for different values of $k$ }
    \label{fig:enter-label}
\end{figure}

\begin{figure}
    \centering
    \includegraphics[width=0.75\linewidth]{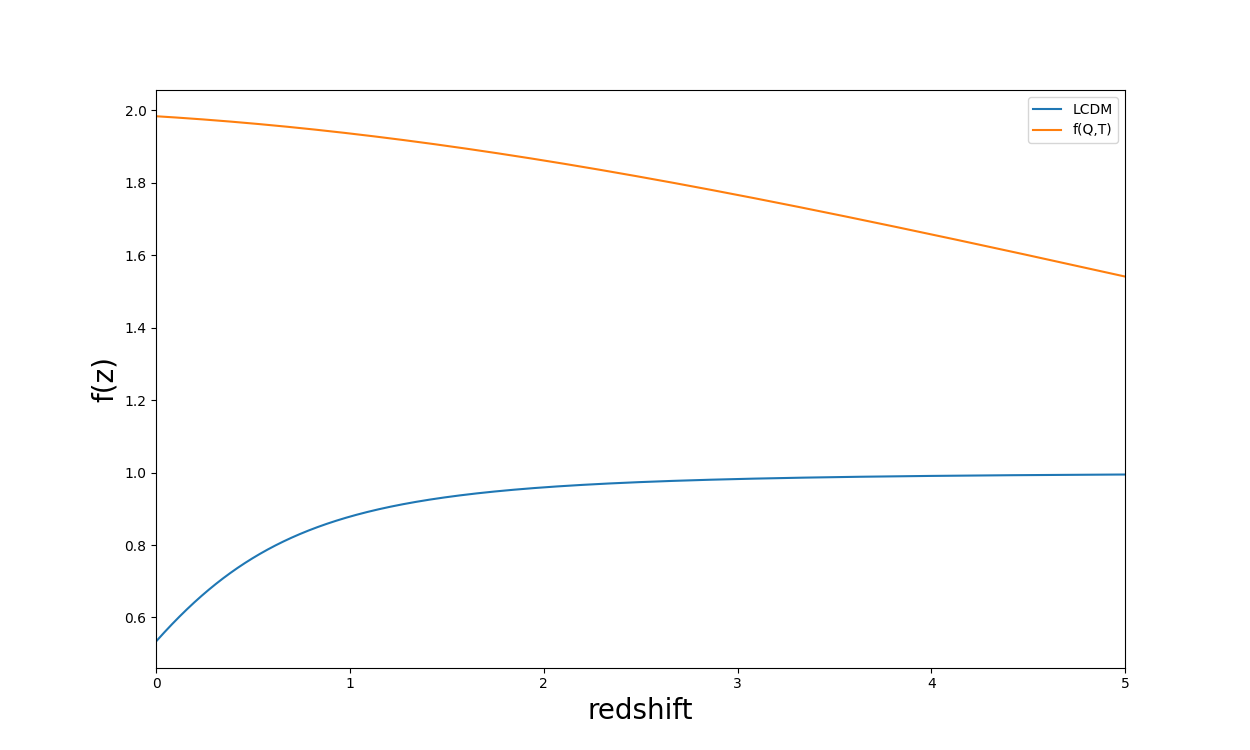}
    \caption{Evolution of the growth factor $f(z)$ with respect to redshift (z) for the case $f(Q,T)=\alpha Q^{n+1}+\beta T$ with $\alpha = -2.5$ , $\beta = -11.5$ and $n= 1$ for different values of $k$ }
    \label{fig:enter-label}
\end{figure}

 \subsection{Holographic energy density}

 The Holographic principle, as an abstract concept, states that all the information contained in a volume of space can be reflected on the horizon surface of that volume \cite{17}.
 When applied on the cosmological level, it was shown that if we assume the vacuum energy of a volume of space, whose surface is of size $L$, cannot exceed the mass of a black of the same size\cite{21,22}, then we find  $L^{3}\rho_{\Lambda} \le  LM^{2}_{P}$  with $\rho_{\Lambda}$ the vacuum energy and UV cut off, $M^{2}_{P}$ being the Planck mass and $L$ the IR cut-off. From these principles, Li \cite{20} was able to develop the Holographic dark energy model with the HDE density taking the following form 
 \begin{equation}
     \rho_{HDE}=\frac{3 C^{2}M^{2}_{Pl}}{L^{2}}
 \end{equation}
 For simplicity, we consider the C parameter to be equal to one and $M^{2}_{Pl}=1$. \\
 
 The choice of $L$ is critical to the form of the HDE model we will be working with.
In this case, just like in \cite{14}, we choose to work with a generalized form of the IR cut-off, as proposed in \cite{23,24} where $L$ is a function of the Hubble parameter, particle horizon, and future horizon. Considering the same special case, we retrieve  : 
\begin{equation}
    L_{IR}=H^{-1}
\end{equation}
Finally, the energy density for HDE is given by 
\begin{equation}
    \rho_{DE}=3L^{-2}=3H^{2}
\end{equation}
We can now proceed to reevaluate the previous model using the HDE dark energy density. \\
In Fig.2, we can see the influence of HDE on the shape of the plot,as we find a cone shaped evolution. While the new energy density is unable to fix the oscillations and divergence issue, it is clear that it was able to decrease from its intensity. We can also clearly see a deviation in the amplitude evolution of the matter density at late times. When implemented with an energy density derived from quantum fluctuations, we retrieve an oscillating but growing baryonic matter density over time in the expanding universe. 
What we can confirm for certain is that, in all cases, the matter density contrast is constant for high redshifts where DE is very subdominant. \\
We conclude that the matter density contrast acts as if non vanishing scalar fields contribute to the total perturbed pressure of baryonic matter density, simulating the forced oscillation standard equations and creating a strong amplitude splash in the generalized $f(Q,T)$ model, then a growing oscillating ampltide of matter density at late-time for the Holographics scenario.
While the Holographic Dark Energy approach to the $f(Q,T)$ gravity showed very promising results in the background framework as seen in \cite{14}, it was unable to reproduce the same level of success when investigated in the context of cosmological perturbation theory. However, it improved the results extracted from the matter density significantly.

\begin{figure}
    \centering
    \includegraphics[width=0.75\linewidth]{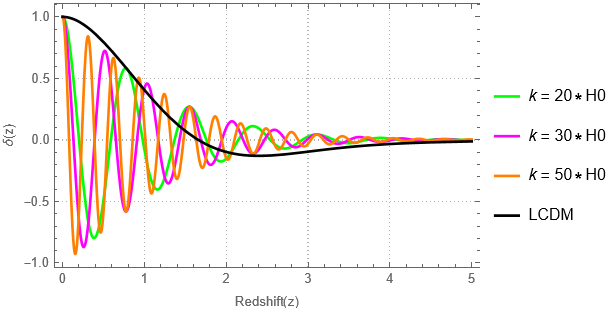}
    \caption{Evolution of density contrast with respect to redshift (z), with HDE for the case $f(Q,T)=\alpha Q^{n+1}+\beta T$ with $\alpha = -2.5$ , $\beta = -11.5$ and $n= 1$ for different values of $k$ }
    \label{fig:enter-label}
\end{figure}

\section{Conclusion}
In this work, we used the theory of cosmological scalar perturbation in $f(Q,T)$, with Q the non-metricity scalar and T the trace of the stress-energy tensor, to find the explicit expression for the matter density equation. Following this study, we observed that the presence of T in the Lagrangian breaks the conservation of the equation, leading to a divergence and the appearance of the hypermomentum. \\
We find the equation for the matter density contrast $\delta$ by deriving the Euler and Continuity equations, and then we assume the sub-Hubble mode and neglect Bardeen's potential time derivative for simplicity. We present the required constraints to be satisfied by this theory to avoid the non-singularity for the non-metricity parameter. \\
Once initial parameters have been established, we find an equation that behaves as a quasi-static approximation of the $f(Q,T)$ gravity, without being actually defined in that limit, and we show that the density contrast obeys a second-order differential equation with strong wave-number dependence in the case of $f(Q,T)=\alpha Q^{n+1} + \beta T$ , the same applies for Holographic dark energy. We then compare our results with the matter density evolution of general relativity and show that the two models, while behaving similarly at high redshifts, evolve in a different manner at late time.
The density contrast evolves as a strong harmonic oscillation for all values of $k$, diverging from General Relativity results around z = 2, as seen in Fig. 1.  \\

We then introduced the Holographic dark energy density into our perturbed models of $f(Q,T)$ in an attempt de explore the effects of quantum entropy on the behavior of perturbation in this specific case of modified gravity. Moreover, the results we obtained allowed us to compare the evolution of density contrast in the $f(Q,T)$ model with the perturbation evolution of other similar models, more specifically the behavior of the perturbed $f(Q)$ model and the $f(R,T)$ model. We were able to conclude that while deviation from GR was expected in this case, the oscillations for the density contrast function appear only when we introduce the stress-energy tensor coupling to our model, which means the evolution of matter density contrast is, in principal, emerging from the behavior of baryonic matter at late-time. We first see an amplitude short-term splash that behaves as probabl frequency resonance, then in the case of Holographic dark energy and the integration of quantum fluctuation, we see an oscillating but constantly growing baryonic matter density at late time, which could be an underlying matter presence only detected at perturbation. The theory might also suffer from strong coupling problems, solutions can be found from looking into degrees of freedom (maybe ghosts). While this subject will be further discussed in future work, it is not within the scope of this paper.\\

The deviation from General Relativity that we see in the $f(Q,T)$ models presented in this paper also gives rise to probable contradictions between our gravity and several observational data sources that can strongly constrain this particular theory. Further investigation using the CMB, power spectrum of galaxy clusters and Planck would provide more information on the disagreement or viability of the model and more context on its behavior regarding the Hubble tension.

\end{document}